\documentclass[reprint,amsmath,amssymb,aps,superscriptaddress, prxquantum]{revtex4-2}

\usepackage{graphicx}
\usepackage{caption}
\usepackage{subcaption}
\usepackage{ragged2e}

\DeclareCaptionJustification{justified}{\justifying}

\usepackage{dcolumn}
\usepackage{bm}
\usepackage{amsmath,amssymb, dsfont}
\usepackage{url}
\usepackage{color}
\usepackage{hyperref}
\usepackage[ruled,vlined]{algorithm2e}

\usepackage{tabularx}
\usepackage{enumitem}
\usepackage{array}

\usepackage{tikz}
\usetikzlibrary{backgrounds,fit,positioning,calc,arrows.meta}
\usepackage{xcolor}
\definecolor{softred}{RGB}{255,210,210}      % fill
\definecolor{softredborder}{RGB}{220,120,120} % optional border
\definecolor{acefill}{RGB}{228,220,255}   % soft amber
\definecolor{aceborder}{RGB}{210, 160, 80}  % optional border
\definecolor{lightamber}{RGB}{255, 236, 200}

\begin{document}

\preprint{APS/123-QED}

\title{Attention-based optimizer for symmetry finding}

\author{Shreya Banerjee}
\email{s.banerjee3@exeter.ac.uk}
\affiliation{Department of Physics and Astronomy, University of Exeter, Stocker Road, Exeter EX4 4QL, United Kingdom}
\affiliation{QuAOS collaboration}

\author{Vinodh Raj Rajagopal Muthu}
\affiliation{Institute for Quantum Computing, University of Waterloo, Waterloo, ON N2L 3G1, Canada}

\author{Charlie Nation}
\affiliation{Department of Physics and Astronomy, University of Exeter, Stocker Road, Exeter EX4 4QL, United Kingdom}
\affiliation{QuAOS collaboration}

\author{Rick P. A. Simon}
\affiliation{Department of Physics and Astronomy, University of Exeter, Stocker Road, Exeter EX4 4QL, United Kingdom}
\affiliation{QuAOS collaboration}

\author{Francesco Martini}
\affiliation{QuAOS collaboration}

\author{Alessandro Ricottone}
\affiliation{QuAOS collaboration}

\author{Federico Cerisola}
\affiliation{Department of Physics and Astronomy, University of Exeter, Stocker Road, Exeter EX4 4QL, United Kingdom}
\affiliation{QuAOS collaboration}

\author{Luca Dellantonio}
\email{l.dellantonio@exeter.ac.uk}
\affiliation{Department of Physics and Astronomy, University of Exeter, Stocker Road, Exeter EX4 4QL, United Kingdom}
\affiliation{QuAOS collaboration}

\begin{abstract}
Finding symmetries is crucial for understanding physical models. In this work, we present an optimization framework that searches Pauli symmetries of Hamiltonians, merging the fields of machine learning with automated symmetry finding. Built on a Set-Transformer architecture, our framework uses self-attention to encode the pairwise and higher-order correlations among the Pauli-Strings. The relations are then decoded as a candidate, which is further optimized with a custom commutation-based objective, and mapped to a symmetry of the input Hamiltonian. We apply our method to random Pauli Hamiltonians, periodic one and two dimensional transverse-field Ising model and the Toric code. We show that for physical Hamiltonians (Ising and Toric), our framework succeeds with near-deterministic probability while providing substantial advantage compared to state-of-the-art strategies. For random Pauli Hamiltonians, we estimate the required computational resources, specifically the number of parallel starts and the number of GPUs, to find a symmetry with high success probability under fixed design specifications.  
\end{abstract}

\maketitle

\section{Introduction}\label{sec:Intro}

Historically, symmetries have always helped to understand and solve complex systems \cite{symm_Noether1918, symm_Gross1996RoleOfSymmetry, symm_wigner}. With the rise of modern computers, it has been possible to study complicated problems directly, without exploiting symmetries \cite{metropolis1953equation, car1985unified, Landau_Binder_2021, exact_diag_1, exact_diag_2}. However, there are still several problems \cite{exact_diag_nonsol1, exact_diag_nonsol2, exact_diag_nonsoln3} that cannot be addressed with brute-force numerical implementations (e.g., exact diagonalisation \cite{banks2022pseudospectral}) alone. Several approximation techniques have been developed to reduce the requirement of computational resources. Among the most successful are tensor-network techniques, such as matrix product states \cite{TN_mps1, TN_mps2, TN_mps3}, projected entangled pair states \cite{TN_mps3, TN_peps1, TN_peps2}, and tree tensor networks \cite{TN_ttn1, TN_ttn2, TN_ttn3}, that have enabled approximate studies of several problems previously thought to be intractable \cite{TN_toughones1, TN_toughones3}. 

Recent developments in machine learning have introduced Probably Approximately Correct methods \cite{algo_pac}. Notable examples include Hamiltonian Neural Networks \cite{ML_HNN1, ML_HNN2} and Lagrangian Neural Networks \cite{ML_LNN1, ML_LNN2}, which enables learning the Hamiltonian and Lagrangian of a system respectively, as well as Graph Neural Networks \cite{ML_GNN1, ML_GNN2}, which have been successfully used to simulate dynamics of complex physical systems. Additionally, tools such as Bayesian optimization \cite{ML_bays1, ML_bays2, TN_toughones2} and reinforcement learning \cite{ML_RL, ML_RL2} have been used to design and optimize variational Ans\"atze for many-body systems.

A common limitation of all these techniques is that they are developed considering specific structural assumptions about the class of problems they aim to model, and their performances degrade when the underlying physical system does not match those assumptions \cite{ML_problem_specific1, ML_problem_specific2, ML_problem_specific3}. As a powerful mitigation, it is possible to rely once more on symmetries to reduce the computational complexity of simulating any general physical system. 

A general, structure-independent algorithm for tapering off qubits was presented in Refs.~\cite{symm_taperingqubits, symm_Luca_min_qrep, van_den_Berg_2020circuit}. The algorithm works with any input Hamiltonian and uses Clifford operations \cite{GottesmanThesis1997, GottesmanHeisenberg1998, Cliff_2} to find a reference frame where all symmetries are stabilized by the computational states. The algorithm then reduces the qubits needed to represent the Hamiltonian, removing redundant ones as well as identifying conserved charges for a block-diagonal representation \cite{symm_Luca_min_qrep}. However, although classically efficient (cubic runtime with respect to the number of qubits), the deterministic guarantee of the algorithm in Refs.~\cite{symm_taperingqubits, symm_Luca_min_qrep, van_den_Berg_2020circuit} realistically comes with a long time-scale for systems with large number of qubits.

In this work, we develop a machine-learning-based optimization framework to identify symmetries. 
Compared to other methods \cite{ML_symm1, ML_symm2, ML_symm3, ML_symm4,  ML_symm5}, ours learns the symmetries directly from any input Hamiltonian.
Furthermore, our approach probabilistically finds Pauli symmetries much faster than the deterministic alternatives in Refs.~\cite{symm_Luca_min_qrep, symm_taperingqubits, van_den_Berg_2020circuit}, with high chances of success and without any prior knowledge about the system or the symmetry. To our knowledge, this is the first time machine learning and artificial intelligence are used to find symmetries directly from an input Hamiltonian, and an important step towards extensions to find other classes of symmetries where no optimal nor deterministic strategies are known.

This article is organized as follows. In Secs.~\ref{sec:background} and \ref{sec:framework} we provide the relevant background and outline our architecture, respectively. A detailed discussion on performances of our approach is provided in Sec.~\ref{sec:results}, where we also present a comparison between our architecture and the deterministic algorithm in Ref.~\cite{symm_Luca_min_qrep}. We conclude in Sec.~\ref{sec:conc} with a summary and future directions.

\section{Background}\label{sec:background}

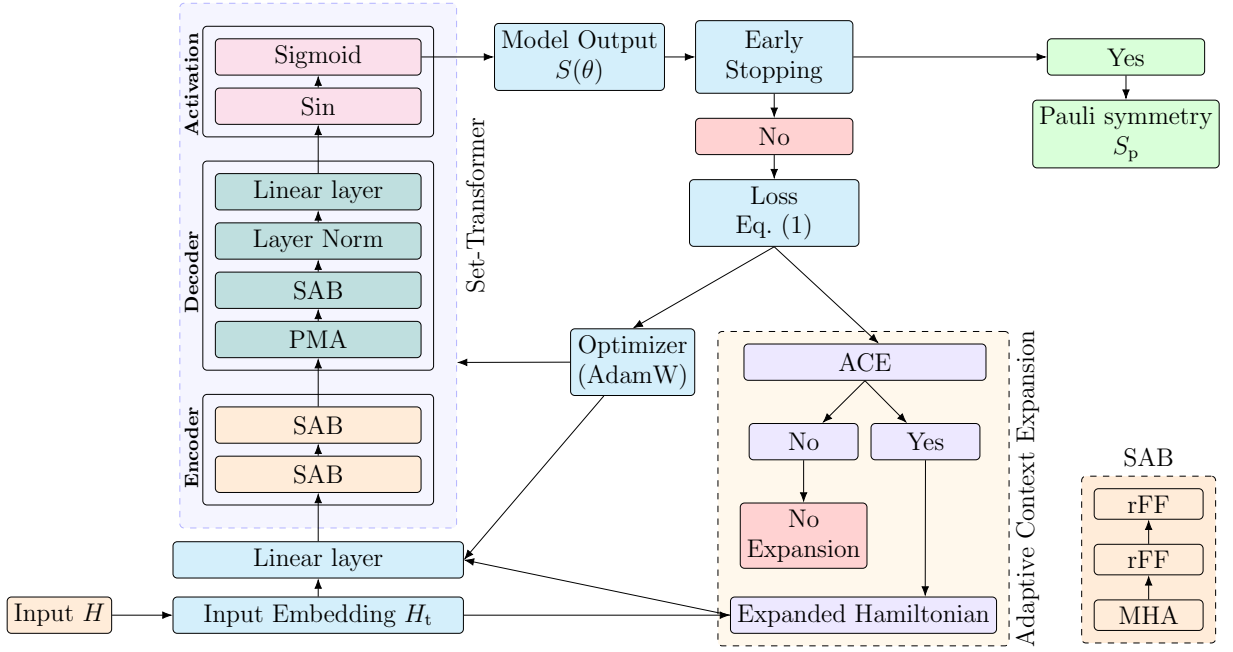
\begin{figure*}[t]
\centering
\resizebox{0.9\textwidth}{!}{%
\begin{tikzpicture}[yscale=1, 
transform shape,
  font=\large,
  >=Latex,
  node distance=6mm and 10mm,
  box/.style={draw, rounded corners=2pt, align=center, minimum height=6mm},
  wbox/.style={box, minimum width=38mm},
  mbox/.style={box, minimum width=34mm},
  sbox/.style={box, minimum width=26mm},
  tiny/.style={draw, rounded corners=2pt, align=center, minimum width=18mm, minimum height=4mm},
  dashedfit/.style={draw, dashed, rounded corners=2pt, inner sep=6mm},
  cproj/.style={box, fill=green!15},
  csab/.style={box, fill=orange!15},
  cdec/.style={box, fill=teal!25},
  cpink/.style={box, fill=magenta!15},
  cblue/.style={box, fill=cyan!15},
  cred/.style={box, fill=softred},
  cace/.style={box, fill=acefill!65},
  copt/.style={draw, rounded corners=2pt, align=center, fill=green!25, minimum width=28mm, minimum height=38mm}
]

% ----------------------------
% Input + Input Projection
% ----------------------------
\node[cblue, box, minimum width=48mm, minimum height=6mm] (inp) {Input Embedding $H_{\rm t}$};
\node[csab, box, minimum width=10mm, minimum height=6mm, left=10mm of inp] (inph) {Input $H$};
\draw[->] (inph) -- (inp);

\node[cblue, box, minimum width=48mm, minimum height=6mm, above=3mm of inp] (inproj) {Linear layer};
\draw[->] (inp) -- (inproj);

% ----------------------------
% Encoder stack (2 SAB)
% ----------------------------
\node[csab, mbox, above=8mm of inproj] (enc1) {SAB};
\node[csab, mbox, above=2mm of enc1] (enc2) {SAB};
\draw[->] (inproj) -- (enc1);
\draw[->] (enc1) -- (enc2);
\node[draw, solid, rounded corners=2pt, inner sep=2mm, fit=(enc1)(enc2)] (encBox) {};
% ----------------------------
% Decoder stack (PMA + SAB)
% ----------------------------
\node[cdec, mbox, above=8mm of enc2] (pma) {PMA};
\node[cdec, mbox, above=2mm of pma] (dsab) {SAB};
\node[cdec, mbox, above=2mm of dsab] (ln) {Layer Norm};
\node[cdec, mbox, above=2mm of ln] (invproj) {Linear layer};

\draw[->] (enc2) -- (pma);
\draw[->] (pma) -- (dsab);
\draw[->] (dsab) -- (ln);
\draw[->] (ln) -- (invproj);

\node[draw, solid, rounded corners=2pt, inner sep=2mm, fit=(pma) (invproj)] (decBox) {};

% ----------------------------
% Activation
% ----------------------------
\node[cpink, mbox, above=8mm of invproj] (sinl) {Sin};
\node[cpink, mbox, above=2mm of sinl] (sigm) {Sigmoid};

\draw[->] (invproj) -- (sinl);
\draw[->] (sinl) -- (sigm);

\node[draw, solid, rounded corners=2pt, inner sep=2mm, fit=(sinl)(sigm)] (actBox) {};

% ----------------------------
% Dashed Set-Transformer container + label
% ----------------------------
\begin{scope}[on background layer]
  \node[dashedfit, fill=blue!4, draw=blue!30, 
  inner sep=5.8mm, 
        fit=(enc1)(sigm)(enc1)(enc2)(pma)(dsab)] (setT) {};
\end{scope}

% label 
\node[rotate=90, font=\large] at ($(setT.east)+(0.3,10mm)$) {Set-Transformer};

% Vertical labels Encoder / Decoder (outside left edge)
\node[rotate=90, font=\small\bfseries] at ($(setT.west)-(-2mm,30mm)$) {Encoder};
\node[rotate=90, font=\small\bfseries] at ($(setT.west)-(-2mm, 1mm)$) {Decoder};
% Vertical label: Activation
\node[rotate=90, font=\small\bfseries] at ($(setT.west)-(-2mm,-30mm)$) {Activation};

% ----------------------------
% Right-side pipeline:
% ----------------------------
\node[cblue, sbox, right=12mm of sigm, minimum height=6mm,] (logits) {Model Output \\ $S(\theta)$};
\node[cblue, sbox, right=5mm of logits, minimum height=12mm,] (es) {Early\\Stopping};
\node[cproj,  sbox, right=32mm of es, minimum height=6mm,] (yes) {Yes};
\node[cproj, sbox, below=4mm of yes, minimum height=10mm] (out) {Pauli symmetry \\ $S_{\rm p}$};

\draw[->] (sigm) -- (logits);
\draw[->] (logits) -- (es);
\draw[->] (es) -- (yes);
\draw[->] (yes) -- (out);
% TikZ style:

\node[cred, sbox, below=4mm of es] (no) {No};
\node[cblue, sbox, below=4mm of no, minimum height=10mm, minimum width= 28mm] (cost) {Loss\\ Eq.~\eqref{eq:loss}};
\node[cace, below=16mm of cost, xshift=15mm, minimum height= 6mm, minimum width= 40mm] (ace) {ACE};

\draw[->] (es) -- (no);
\draw[->] (no) -- (cost);
\draw[->] (cost.south) -- ($(ace.north)+(2.3mm,0)$);

% optimizer
\node[cblue, mbox, left=8mm of ace, minimum height=10mm, minimum width=20mm] (opt)
{Optimizer\\ (AdamW)};
\draw[->] (cost.south) -- (opt.north);

% Ace blocks
\node[cace, below=7mm of ace, xshift=10mm, minimum height= 6mm, minimum width= 18mm] (acey) {Yes};
\node[cace, left=2mm of acey, minimum height= 6mm, minimum width= 18mm] (acen) {No};
\node[cred, below=7mm of acen, minimum height= 6mm, minimum width= 18mm] (dn) {No \\ Expansion};
% \node[cblue, below=5mm of acey, minimum height= 6mm, minimum width= 18mm] (dn) {Expand\\ Context};

\node[cace, box, minimum width=40mm, minimum height=6mm, right=44mm of inp] (expcont) {Expanded Hamiltonian};

\draw[->] (ace.south) -- (acen);
\draw[->] (ace.south) -- (acey);
\draw[->] (acen) -- (dn);
\draw[->] (acey.south) -- ($(expcont.north)+(10.2mm,0)$);
\draw[->] (inp.east) -- ($(expcont.west)+(0.0, 0.0mm)$);
% ----------------------------
% Feedback curves into Input Projection Layer 
\draw[->] ($(expcont.west)+(0.0, 0.0mm)$) -- (inproj.east);
\draw[->] (opt) --($(inproj.east)+(0.0, 0.0)$);
\draw[->] (opt.west) -- ($(setT.east)+(0.0,-16.0mm)$);

% % ace block coloring 
\begin{scope}[on background layer]
  \node[dashedfit, inner sep=2mm,  
  fill=lightamber!35,fit=(ace)(expcont)] (aceb) {};
\end{scope}

% label 
\node[rotate=90, font=\large] at ($(aceb.east)+(0.3,0.1mm)$) {Adaptive Context Expansion};

% ----------------------------
% SAB inset (below the output block)
% ----------------------------
\node[tiny, right=16mm of expcont](mha){MHA};
\node[tiny, above=4mm of mha] (rff1) {rFF};
\node[tiny, above=4mm of rff1] (rff2) {rFF};

\draw[->] (mha) -- (rff1);
\draw[->] (rff1) -- (rff2);

\begin{scope}[on background layer]
  \node[dashedfit, inner sep=2mm,  
  fill=orange!15,fit=(mha)(rff2)] (sabInset) {};
\end{scope}

\node[font=\large] at ($(sabInset.north)+(0,3mm)$){SAB};
% \node[rotate=90, font=\small] at ($(sabInset.east)+(0.3,3mm)$) {SAB};
\end{tikzpicture}
}
\caption{
Schematic diagram for the attention-based optimization framework to find a Pauli symmetry $S_{\rm p}$ of the input Hamiltonian $H$. $H$ is first embedded into its tableau representation $H_{\rm t}$ \cite{symm_PRL, symm_companion} in the input embedding layer, and then is passed through a  linear row-wise projection layer \cite{NLP_ST_lee2018transformer, NLP_latent_dim}. The underlying machine-learning model is a Set-Transformer \cite{NLP_ST_lee2018transformer, NLP_T_vaswani2017attention} (light-blue box with dashed border in left), consisting of Encoder-Decoder stacks and an activation module. The encoder (light-orange) contains stacked Set Attention Blocks (SAB) \cite{NLP_ST_lee2018transformer} (two in the figure). The Decoder (dark green) includes a Pooling Multi-head Attention layer (PMA) \cite{NLP_PMA1}, followed by a SAB, layer-norm \cite{NLP_layernorm1}, and a linear layer. The Activation module (pink) contains a Sin and a Sigmoid layers \cite{NLP_sigmoid1, NLP_sigmoid2}. The rest of the optimization framework is shown on the right-side of the Set-Transformer, with an early-stopping decision flow, and the Adaptive Context Expansion module (light yellow box with dashed black border). On the lower right, the modules inside a SAB is presented (light-orange box with black dashed border). The SAB layer contains a Multi-Head Attention module (MHA) \cite{NLP_T_vaswani2017attention} followed by two row-wise Feed-Forward (rFF) \cite{NLP_rFF1} layers.
}
\label{fig:framework}
\end{figure*}

In this work, we consider input Hamiltonians $H = \sum_{i=1}^N P_i$ written as a sum of $N$ $n_{\rm q}$-qubit Pauli-Strings $P_i$. Each $P_i$ is a unique tensor product of $n_{\rm q}$ single-qubit Pauli operators $\{ \mathds{I}, X, Y, Z \}$. As introduced above, our aim is to find a Pauli symmetry $S_{\rm p}$ of $H$, i.e. a Pauli-String that commutes with all $P_i$ within $H$. For a system with $n_{\rm q}$ qubits, the operator $H$ is a $2^{n_{\rm q}} \times 2^{n_{\rm q}}$ matrix. Thus, to represent $H$ and its constituent $P_i$ efficiently, we employ the symplectic formalism \cite{symm_Luca_min_qrep, dehaene_moor_symplectic1,GottesmanThesis1997, GottesmanHeisenberg1998, tehral_symplectic2, Cliff_2}, which dictates that an $n_{\rm q}$-qubit Pauli-String can be written as a binary vector with $2n_{\rm q}$ entries. The resulting $N \times 2 n_{\rm q}$ tableau representation of $H$ is not unique, as any possible permutation of the rows represents the same Hamiltonian. 

\iffalse
As an example, we consider a $3$-qubit Hamiltonian, \[H = IZI + ZIX+ XXY.\] in its tableau representation, $H$ can be written as (considering $Y=XZ$), 
\[ H_{\rm t} = 
\left(
\begin{array}{ccccccc}
0 & 0 & 0 & \lvert & 0 & 1 & 0 \\
0 & 0 & 1 & \lvert & 1 & 0 & 0 \\
1 & 1 & 1 & \lvert & 0 & 0 & 1
\end{array}
\right).
\]
Considering a permutation of rows $1$ and $3$ in $H_{\rm t}$, 
\[ H_{\rm {s^{\prime}}} = 
\left(
\begin{array}{ccccccc}
1 & 1 & 1 & \lvert & 0 & 0 & 1 \\
0 & 0 & 1 & \lvert & 1 & 0 & 0 \\
0 & 0 & 0 & \lvert & 0 & 1 & 0 \\
\end{array}
\right).
\]
However, the Hamiltonian $H_{\rm {s^{\prime}}}$ encodes is, \[XXY + ZIX + IZI = H.\]
Subsequently, $XIY \cong \left( 1  0  1 \lvert 0  0 1 \right)$ is one of the Pauli symmetries of $H$ that commutes with both $H_{\rm t}$ and $H_{\rm {s^{\prime}}}$. This indicates, permuting the position of the Pauli-Strings in the tableau representation of $H$ does not impact its symmetries.
\fi

Therefore, the problem of finding $S_{\rm p}$ is intrinsically permutation-invariant. 
As a consequence, an automated architecture that finds $S_{\rm p}$ must preserve the permutation-invariance of its input $H$. Introduced in Ref.~\cite{NLP_ST_lee2018transformer}, a Set-Transformer takes an unordered set as its input and encodes the long-range interactions among the elements of the set into its output, while maintaining the permutation-invariance of the input. This makes the Set-Transformer a suitable choice as the underlying model to detect the Pauli symmetries of $H$. 

Furthermore, since $S_{\rm p}$ commutes with each $P_i$ in $H$,  we can exploit this \textit{known} correlation between $S_{\rm p}$ and the elements of $H$ to design custom loss functions for our optimization framework. Let us first introduce the relevant quantities. From the Hamiltonian $H$, its tableau representation \cite{symm_PRL, symm_companion} is $H_{\rm t}$. The output of the machine-learning model (see below and Fig.~\ref{fig:framework}), that depends on the learnable model parameters $\theta$, is $S(\theta)$. $S(\theta)$ is a $2n_{\rm q}$ vector that encodes the desired symmetry $S_{\rm p}$. Finally, 
\begin{equation*}
    J =
    \left(
    \begin{array}{ccccccc}
    0 & \mathds{I} \\
    \mathds{I} & 0
    \end{array}
    \right)
\end{equation*}
is the $2n_{\rm q} \times 2n_{\rm q}$ canonical tableau form on the vector space $\mathds{F}_2^{2n}$ \cite{GottesmanThesis1997, sympleq_Haah_2017}, that serve to efficiently calculate the commutator between operators in their tableau forms.

The primary loss function of our architecture is $\mathcal{C}_{\text{com}}(\theta)$, which exploits the commutation relation between the constituent Pauli-Strings of $H$ and $S_{\rm p}$. Additionally, we use three penalties $\mathcal{C}_{\text{zp}}$, $\mathcal{C}_{\text{bin}}$ and $\mathcal{C}_{\text{lin}}$ to ensure that a non-trivial Pauli symmetry is efficiently found. The loss function $\mathcal{C}$ is then 
\begin{equation*}
    \mathcal{C} 
    = 
    \alpha_{\text{com}} \mathcal{C}_{\text{com}} 
    + 
    \alpha_{\text{zp}} \mathcal{C}_{\text{zp}} 
    + 
    \alpha_{\text{bin}} \mathcal{C}_{\text{bin}} 
    + 
    \alpha_{\text{lin}} \mathcal{C}_{\text{lin}},
\end{equation*}
with $\alpha_{\textunderscore}$ scalars that prioritize different components of $\mathcal{C}$ (see Appendix~\ref{appen:loss}) and
\begin{subequations}\label{eq:loss}
    \begin{align}
        \mathcal{C}_{\text{com}}(\theta) 
        &= 
        \sum_{k=1}^{N}\sin^2
        \left(
        \frac{\pi}{2}[ (H_{\rm t}J)_k S(\theta)^{T} ]
        \right)
        , \label{eq:loss_com}
        \\
        \mathcal{C}_{\text{zp}}(\theta) 
        &= 
        \left(
        \sum_{i=1}^{2n_{\rm q}}S(\theta)_i
        \right)^{-2}
        , \label{eq:loss_zp}
        \\
        \mathcal{C}_{\text{bin}}(\theta)
        &=
        \sum_{i=1}^{2n_{\rm q}}
        S(\theta)_i
        \left(
        1-S(\theta)_i
        \right)
        , \label{eq:loss_bin}
        \\
        \mathcal{C}_{\text{lin}}(\theta) 
        &=
        \sum_{k=1}^{N} \left\lvert 
        (H_{\rm t}J)_k S(\theta)^{T}
        - \text{ne}
        \left(
        \sum_{l} (H_{\rm t}J)_{kl}
        \right)
        \right\rvert
        , \label{eq:loss_lin}
    \end{align}
\end{subequations}
where $\text{ne} \left( \cdot \right)$ returns the nearest even integer of the number within.

The main loss function $\mathcal{C}_{\text{com}}(\theta)$ in Eq.~\eqref{eq:loss_com} is a differentiable extension of the commutator $[P_i,S_{\rm p}] = (H J)_i S(\theta_{\rm p})^{T} \mod{2}$, where the subscript `p' in $\theta_{\rm p}$ indicates that $S(\theta_{\rm p})$ is the tableau of $S_{\rm p}$. We use $\sin^2(\frac{\pi}{2}x)$ as a differentiable proxy for $x\mod{2}$. For integer $x$, this proxy reproduces  $x\mod{2}$ exactly, while providing smooth gradients when $x$ is relaxed to be continuous during optimization. The downside of using $\sin^2(\frac{\pi}{2}x)$ in Eq.~\eqref{eq:loss_com} is that it is highly nonlinear and multi-modal, and optimization can struggle near a local minimum. The penalties below and specifically $\mathcal{C}_{\text{lin}}(\theta)$ in Eq.~\eqref{eq:loss_lin} mitigate this problem.

The first penalty $\mathcal{C}_{\text{zp}}(\theta)$ in Eq.~\eqref{eq:loss_zp} is added to avoid finding the Identity $S(\theta)_i = 0$ for all $i=1,\dots, 2n_{\rm q}$, the trivial Pauli symmetry. The second penalty $\mathcal{C}_{\text{bin}}(\theta)$ in Eq.~\eqref{eq:loss_bin} forces outputs $S(\theta)$ of our architecture towards binary values $0$ and $1$. Indeed, as we use the classically efficient tableau representation of the Pauli-Strings, the desired output $S(\theta_{\rm p})$ needs to be a vector of binary numbers. Finally, the third penalty $\mathcal{C}_{\text{lin}}(\theta)$ in Eq.~\eqref{eq:loss_lin} is a linearity regularizer. It assists the model during early optimization phases by favoring symmetries that anti-commute limited numbers of times with the Hamiltonian. We discuss the loss functions in more detail in Appendix~\ref{appen:loss}.

Finding a Pauli symmetry has also an implicit dependence on the inherent algebraic properties of $H$. Calling $\mathcal{R}$ the rank of $H_{\rm t}$, the number of Pauli-Strings that generate the set of all possible Pauli symmetries of $H$ is $2n_{\rm q}-\mathcal{R}$ \cite{symm_Luca_min_qrep}. As $\mathcal{R}$ increases, the fraction of Pauli-strings that are symmetries of $H$ decreases as $1/2^{\mathcal{R}}$, given that the total number of possible Pauli-Strings on $n_{\rm q}$-qubits is $2^{2n_{\rm q}}$. For an optimization framework this means that as $\mathcal{R}$ increases the loss-landscape becomes skewed with exponentially many non-solutions compared to solutions. Despite this not being a problem for the physical examples considered below, it is for random Hamiltonians, see Sec.~\ref{sec:results}. As a mitigation, it is possible to increase the context provided to the architecture \cite{leviathan2025selective}. As explained in Sec.~\ref{sec:framework}, this is what the Adaptive Context Expansion (ACE) does.

\section{Framework}\label{sec:framework}

Our framework is motivated by an analogy between Natural Language Processing (NLP) tasks \cite{NLP1, NLP2, NLP_T_vaswani2017attention}, and a Hamiltonian $H$. Analogous to the unique tokens in the vocabulary of a natural language, each $P_i$ can be thought of as a word in a language that has $4^{n_{\rm q}}$ distinct words. The Hamiltonian $H$ then becomes a collection of such words and its symmetries reflect a global relation among them. This  naturally motivates a machine-learning architecture, developed to extract long-range relations across the words of an input text. 

Despite the similarities, the distinct feature that separates our problem from NLP is that all $4^{n_{\rm q}}$ words could be Pauli-Strings of some Hamiltonian. In contrast, natural languages only consider a subset of all possible combinations of letters. This implies that the number of candidates for the output Pauli symmetry is exponential in $n_{\rm q}$, making an input-output dataset and training the Set-Transformer expensive and inefficient. 
As a solution, we utilize the commutation relation between the Hamiltonian $H$ and its symmetry $S_{\rm p}$ to formulate the problem of finding $S_{\rm p}$ as a minimization objective.

To extract the commutation relation in the form of $S_{\rm p}$, the Set-Transformer utilizes \textit{self-attention} \cite{NLP_T_vaswani2017attention}. In self-attention, each Pauli-String in $H$ is mapped to three distinct learnable vectors: a \textit{query} vector $q_i$, a \textit{key} vector $k_i$, and a \textit{value} vector $v_i$. Stacking these vectors together, three learnable matrix representations of $H$ are obtained: a query $Q$, key $K$, and value $V$. The attention scores, which qualitatively encode the relevance between two elements of the input Hamiltonian, is then computed as  
\cite{NLP_T_vaswani2017attention}, 
\[\text{Attention}(Q, K, V) = \text{Softmax}\left(\frac{QK^T}{\sqrt{d_k}}\right)V,\] where $d_k$ is the dimension of the key vectors. Computing the self-attention for $H$ thus produces an output matrix that reflects the learned correlations among the Pauli-Strings in $H$. The Set-Transformer later projects these encoded correlations into the form of a Pauli-String (see below), which, after the optimization, becomes the Pauli symmetry $S_{\rm p}$ of $H$.

A schematic diagram of the entire optimization flow is provided in Fig.~\ref{fig:framework}, with all the building blocks, i.e., the Input Embedding and Projection, the Set-Transformer and the ACE modules described below.
% }

\noindent\textbf{Input Embedding and Projection:} In NLP, the words are embedded into unique, continuous, higher-dimensional learnable tokens prior to being processed by the underlying model \cite{NLP_word_embedding}. Analogous to words in a text, the tokens are the unique Pauli-Strings in $H$. As discussed above, the tableau provides a meaningful representation in $2n_{\rm q}$ dimensions \cite{symm_PRL, symm_companion}. We choose to use it as the primary Input Embedding (blue in Fig.~\ref{fig:framework}) of the Pauli-Strings. 
To preserve the permutation-invariance of the problem we avoid position embedding, as typical for a Set-Transformer \cite{NLP_ST_lee2018transformer}. 
After the input embedding, to incorporate the learned nature of the embeddings, we use a linear layer with tunable weights. This layer row-wise projects the binary vector representing each $P_i$ to a continuous, learnable vector, while the row-wise projection maintains the permutation-equivariant nature of the architecture \cite{NLP_ST_lee2018transformer}. This layer can also be used to project the input tokens to a latent embedding dimension $n_{\text{emb}}$ \cite{NLP_latent_dim}, to further increase the learnability of the model. The output of this linear layer in Fig.~\ref{fig:framework} is thus an $N \times n_{\text{emb}}$ matrix, which then acts as the input to the Set-Transformer.

\noindent\textbf{Set-Transformer:} The Set-Transformer processes the embedded input Hamiltonian and extract the correlations among its Pauli-Strings in the form of another Pauli-String. A schematic diagram is provided in Fig.~\ref{fig:framework} (light-blue box with dashed border). It is composed of an encoder stack (light-orange), followed by a decoder stack (green-blue) \cite{NLP_T_vaswani2017attention, NLP_ST_lee2018transformer} and an activation module (pink). Each layer of the Set-Transformer is permutation-equivariant \cite{NLP_ST_lee2018transformer}, maintaining the permutation invariant nature of the problem (see above).

The encoder stack (light-orange in Fig.~\ref{fig:framework}) in the transformer contains Set Attention Blocks (SAB, see figure) \cite{NLP_ST_lee2018transformer}, which are built using Multi-Head Attention layers (MHA) \cite{NLP_T_vaswani2017attention}, followed by fully connected row-wise Feed-Forward (rFF) layers \cite{NLP_ST_lee2018transformer, NLP_rFF1}. Through self-attention \cite{NLP_T_vaswani2017attention}, the SAB layers in the encoder capture the pairwise correlations among the Pauli-Strings of the Hamiltonian $H$. 
Stacking multiple SABs allows the encoder to capture progressively higher-order correlations, producing an output that reflects the learned interactions among all Pauli-Strings.

The decoder stack (green in Fig.~\ref{fig:framework}) projects the relations learned by the SABs in the encoder into a single Pauli-String (i.e., $2n_{\rm q}$-dimensional binary vector), which will then become the desired Pauli symmetry. The stack contains a Pooling Multi-head Attention (PMA) module \cite{NLP_PMA1, NLP_ST_lee2018transformer}, followed by a SAB, a layer norm \cite{NLP_T_vaswani2017attention, NLP_layernorm1, NLP_tfixup} and a linear layer that performs a row-wise projection \cite{NLP_latent_dim}. PMA also implements multi-head attention to extract information from the encoder output. The layer uses the output matrix of the encoder stack and a learnable seed vector with $n_{\text{emb}}$ entries to aggregate the learned correlations of the encoder as one vector of length $n_{\text{emb}}$. Inspired by Ref.~\cite{NLP_ST_lee2018transformer}, a SAB following the PMA allows the output vector to better represent the correlations learned by the transformer. 

After the SAB, we implement a layer normalization block within the decoder, which rescales the elements of the output vector to help stabilize gradient updates during training \cite{NLP_layernorm1, NLP_tfixup}. The output of this block is then passed through a linear layer that transforms the length of the query vector from $n_{\text{emb}}$ to $2n_{\rm q}$. This dimensional conversion of the output vector is done to project the learned correlations to a vector with the dimension $2n_{\rm q}$ of the Pauli-String tableau.

Finally, the output of the decoder stack is passed through an activation module (pink in Fig.~\ref{fig:framework}) consisting of a Sin and a learnable (see below) Sigmoid layer. The activation module is motivated by the binary nature of the entries in the tableau representation of a Pauli-String. The Sin layer projects the output of the transformer to $[-1, +1]$, whereas the Sigmoid layer converts the individual continuous-valued entries of the model output to their closest approximations of $0$ and $1$. To do so optimally, we let the architecture learn the best possible threshold and temperature of the Sigmoid  \cite{NLP_sigmoid1, NLP_sigmoid2}. A smaller (higher) value of the temperature ensures that assignment of the variables from $[-1, +1]$ to $[0, 1]$ happens on a steeper (smoother) curve, whereas the threshold determines the transition point. An `optimal choice' of the Sigmoid temperature in our context thus refers to a transition Sigmoid curve that is not too steep, which could prevent an appropriate regression, nor too smooth, which could cause the model to yield a vector with elements far from the extrema of the range $[-1, 1]$ prior to the Sigmoid. Additionally, a learnable threshold ensures a transition point adaptable to the input, suitable for an optimization framework.

The output of the Set-Transformer is thus a $2n_{\rm q}$-dimensional vector with continuous valued entries approximating binary numbers. As optimization progresses, this vector reflects the the global correlations among the Pauli-Strings in the input Hamiltonian $H$, and eventually is expected to turn into a Pauli symmetry of $H$.

\noindent{\textbf{Adaptive Context Expansion}}: In NLP, the unique embedded tokens in the input are the `context' \cite{leviathan2025selective} available to a transformer, which in turn extracts the global correlations among them. In our architecture, the attention layers in both encoder and decoder stacks capture the pairwise and higher order correlations among the Pauli-Strings of the input Hamiltonian. This implies that the number of Pauli-Strings in the input Hamiltonian corresponds to the size of the context available to the Set-Transformer. As mentioned before, with an increasing rank $\mathcal{R}$ of a Hamiltonian, the number of symmetries decreases exponentially. Ref.~\cite{leviathan2025selective} shows that self-attention benefits from diverse and informative context relevant to the learning objective. Thus, to help the optimization progress in a regime where there are exponentially many non-solutions compared to the solutions, we introduce a helper module, the Adaptive Context Expansion (ACE, light yellow in Fig.~\ref{fig:framework}). ACE synthetically increase the context available to the transformer by exploiting the fact that the product $P_i P_j$ of two Pauli-Strings in $H$ commutes with every Pauli symmetry $S_{\rm p}$ of $H$. 

After a user-specified number of runs, ACE checks whether the optimizer is stuck in a local minimum. This is done by identifying oscillating behaviours in the loss functions in Eq.~\eqref{eq:loss}, which indicates that the optimizer is stuck in a corner of the parameter space $\theta$. ACE then expands the context by adding a chosen number of Pauli-Strings to $H$ and updating the input to the model. This effectively provides a kick to the optimizer that allows it to escape the loop and keep searching elsewhere. To prevent increasing the number of Pauli-Strings arbitrarily, ACE is adaptive -- see Appendix~\ref{appen:ace} for a pseudocode and other details.

\noindent{\textbf{Optimization Flow}}: With all modules (Input Embedding and Projection, Set-Transformer and ACE) introduced, we can describe the optimization flow as presented in Fig.~\ref{fig:framework}. The framework is designed to be an anytime algorithm \cite{algo_anytime}, with early stopping enabled (see Appendix~\ref{appen:mlspecs}).
At every iteration, the output from the Set-Transformer is passed to check if it satisfies the early stopping conditions; namely, whether a symmetry is found (which is done efficiently by commuting it with the Hamiltonian). If it does, the optimization stops, and the output is accepted. If the check fails, the loss functions in Eq.~\eqref{eq:loss} are computed and passed on to the underlying gradient-based optimizer, and the model parameters are updated.

Additionally, the computed objectives are also supplied to the ACE module, and depending on their behavior (see above) ACE iteratively expands the context, supporting convergence of our algorithm.

\section{Results}\label{sec:results}
\begin{figure*}[!hbt]
    \centering
    \begin{subfigure}[t]{0.329775\textwidth}
        \centering
        \caption{Random Hamiltonians.}
\label{fig:results_random}
\fbox{
\includegraphics[height=11.5cm]{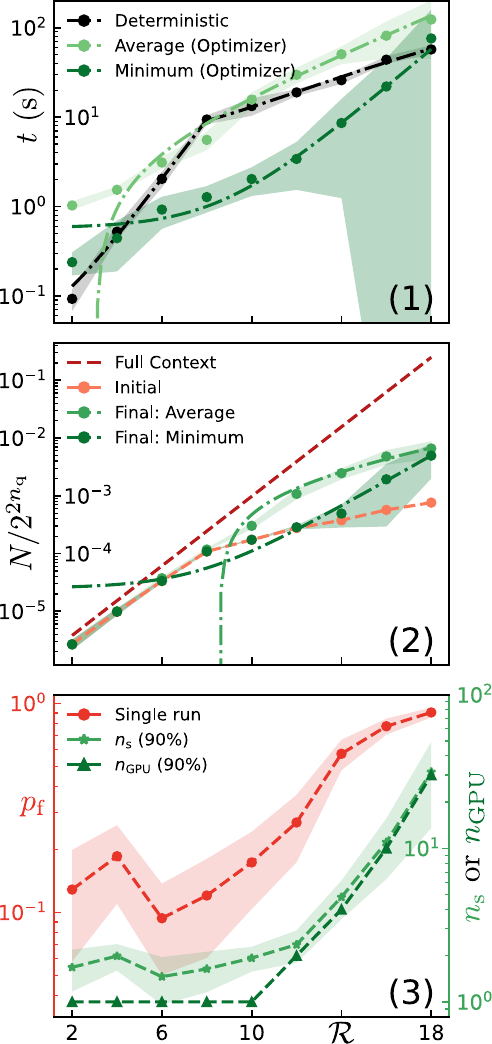}
}
    \end{subfigure}
    \hfill
    \begin{subfigure}[t]{0.329775\textwidth}
        \centering
        \caption{Periodic $1$-D Ising Chain.}
\label{fig:results_ising}
\fbox{
\includegraphics[height=11.5cm]{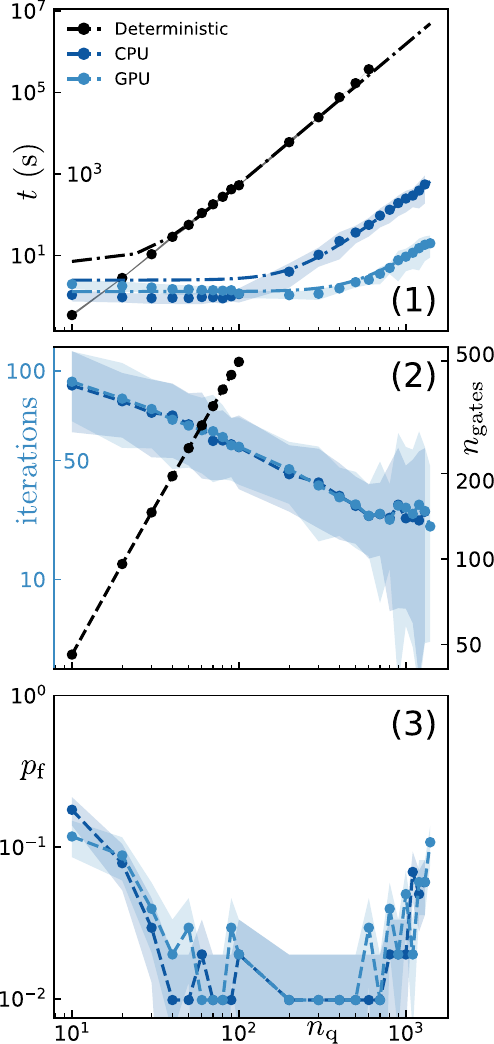}
} 
    \end{subfigure}
    \hfill
    \begin{subfigure}[t]{0.329775\textwidth}
        \centering
        \caption{Physical Hamiltonians.}    \label{fig:results_toric}
        \fbox{
        \includegraphics[height=11.5cm]{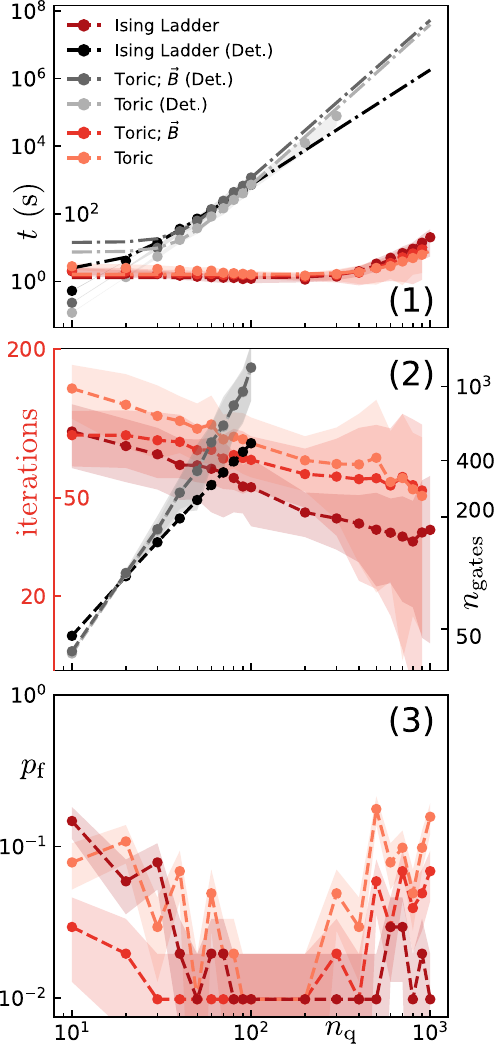}
        }
    \end{subfigure}
    \caption{
    (a): Benchmark on random Hamiltonians with $10$ qubits and increasing $\mathcal{R}$. a(1) shows the average (light-green) and minimum (dark-green) times taken to find a symmetry. a(2) showcases the performance of the ACE module. We plot the average (light-green) and minimum (dark-green) contexts $N$ needed to find a symmetry, rescaled by the full context (red dashed line) $2^{\mathcal{R}}$. The orange dashed line reports the initial number of Pauli-Strings $\approx\mathcal{R}^2$. a(3) presents the probability of failure $p_{\rm f}$ (red) of the attention-based framework and the number of parallel starts $n_{\rm s}$ (green stars) required by the optimizer to achieve a $90 \%$ success rate, as well as the number of GPUs $n_{\rm GPU}$ (green triangles) to enable those starts. (b): Performance of the CPU (dark-blue) and GPU (light-blue implementations of our framework as well as the deterministic approach (black), with the periodic $1$-D transverse-field Ising Hamiltonian as a model \cite{Ising_chain}. b(1) present the time to find a Pauli symmetry, corresponding to the iterations (number of gates $n_{\rm gates}$) shown in b(2) for our framework (the deterministic algorithm). In b(3), we report the failure probability $p_{\rm f}$. (c): Performance of the GPU implementation of our framework [deterministic algorithm] applied to the $2$-D TFI on a ladder \cite{Ising_ladder} (red) [(black)], and Toric code \cite{toric_ham, toric2} with (dark-orange) [(dark-grey)] and without (light-orange) [(light-grey)] a magnetic field $\vec{B}$. c(1) shows the required time, c(2) the number of iterations (our approach) and gates $n_{\rm gates}$ (deterministic algorithm), and c(3) the failure rates $p_{\rm f}$ of the optimizer. In all plots, dots are averages, shadows their corresponding standard deviations, dashed (dash-dotted) lines connect (are fit of) the points. Bayesian statistics \cite{Bays_pf2, Bays_pf3, simon2025enhanced, simon2026error} is employed in a(3), b(3), and c(3); in all other panels we used sample average and variance \cite{shlosberg2023adaptive}. For specification of the GPU and CPU used, see Appendix~\ref{appen:mlspecs}.
    }
    \label{fig:results}
\end{figure*}

In this section, we benchmark our algorithm and discuss the numerical results presented in Fig.~\ref{fig:results}.

\subsection{Random Hamiltonians}\label{ssec:res_rand}

Fig.~\ref{fig:results_random} shows the performance of our architecture against randomly generated Hamiltonians with varying numbers of Pauli symmetries. As explained in Sec.~\ref{sec:background}, this number determines the rank $\mathcal{R}$, which is reported on the $x$ axis for all panels a(1-3). By increasing $\mathcal{R}$, the ratio of Pauli symmetries of a Hamiltonian to non-symmetries decreases exponentially, making the convergence of our framework to a desired output more challenging. We plot the time required to find a symmetry in panel a(1), the context ratio (see below) in a(2), and the failure probability and resource counts in a(3). To systemically examine the performance of the framework, we restrict ourselves to $10$-qubit Hamiltonians and compare the results with the deterministic algorithm in Refs.~\cite{symm_Luca_min_qrep, sympleq} (in black). 

For each rank $\mathcal{R}$, we average over $100$ different random Hamiltonians, each consisting of $N \approx 2\mathcal{R}^2$ Pauli-Strings. Initiating $25$ independent starts (i.e., random initialization of the frameworks' parameters; for details, see Appendix~\ref{appen:mlspecs}), the attention-based optimizer was allowed to run for a fixed number of $5000$, $10000$ and $12000$ iterations per start at $\mathcal{R} \leq 14$, $\mathcal{R} = 16$ and $\mathcal{R} = 18$, respectively. At each iteration, we check whether a symmetry is found and set an early stopping condition in case it was. The fastest [average] time to find a symmetry out of all successful runs determines the minimal [average] time, [light] green in panels a(1,2). The device and design specifications for our simulations are provided in Appendix~\ref{appen:mlspecs}. 
 
Fig.~\ref{fig:results_random}(1) shows that, aided with multiple independent starts and the ACE module introduced in Sec.~\ref{sec:framework}, the attention-based optimizer finds a Pauli symmetry faster than the deterministic algorithm \cite{symm_Luca_min_qrep} for ranks $4$ to $16$, and in a similar time for ranks $2$ and $18$. While the average time taken by the optimizer to find a symmetry (light-green in the figure) is almost always higher than the deterministic algorithm (black), the minimum time (dark-green) is lower for most values of $\mathcal{R}$. 

Dash-dotted lines in Fig.~\ref{fig:results_random} are fits of numerical data. The minimal [average] time taken by our optimizer to find a Pauli symmetry scales as $O(2^{0.705 \mathcal{R}})$ [$O(2^{0.353 \mathcal{R}})$], see the dark-green [light-green] line in panel a(1). The scaling for the minimal time can be improved, at a higher computational cost, by increasing the number of starts (see below). In comparison, the time complexity of the deterministic algorithm (black line) increases as $O(2^{\mathcal{R}})$ up to rank $\mathcal{R} \leq 8$, and then polynomially as $O(\mathcal{R}^{3})$.  When $\mathcal{R} > 8$, this is in agreement with the literature \cite{symm_Luca_min_qrep}. For lower ranks, the different scaling behavior can be ascribed to the number of Pauli-Strings being less than twice the number of qubit. Ref.~\cite{symm_Luca_min_qrep} assumes $N \gg 2n_{\rm q}$ when determining the asymptotic scaling. For our test-case Hamiltonians with $10$ qubits and $N \approx 2\mathcal{R}^2$ Pauli-Strings, $N \geq 2n_{\rm q}$ is satisfied when $\mathcal{R} > 4$, leading to an expected shift in scaling above rank $4$.

We highlight that the deterministic algorithm \cite{symm_Luca_min_qrep} finds all the generators of Pauli symmetries, compared to our framework, which only gets one per independent run. In this work, we are more interested in finding a symmetry as fast as possible. The problem of finding others can be turned into recursive instances, each with a reduced Hamiltonian that has less qubits than the original one \cite{symm_Luca_min_qrep}. Additionally, we note that different runs (starts) of attention-based optimizer generally yield different symmetries of $H$. With sufficiently many parallel starts, it is then possible to retrieve all generators of the Pauli symmetries of $H$ within a time-frame comparable to the algorithm in Ref.~\cite{symm_Luca_min_qrep}. 

In Fig.~\ref{fig:results_random}(2) we show the impact of the ACE module described in Sec.~\ref{sec:framework} on the performance of the optimizer. The figure shows ACE's expansion of the context for the instances where the optimizer has successfully found a symmetry. To quantify this expansion, we define the context ratio of a Hamiltonian $H$ with $n_{\rm q}$-qubits as $N/2^{2n_{\rm q}}$, where $N$, the number of Pauli-Strings in $H$, can be increased when ACE is employed. The context ratio of the initial input Hamiltonians (before ACE) is in orange, while the ones of the Hamiltonians for which a symmetry is found are in green. The red dashed line indicates the full context, i.e. the upper bound $2^{\mathcal{R}-2n_{\rm q}}$ given by all possible Pauli-Strings in a Hamiltonian with $n_{\rm q}$ qubits and rank $\mathcal{R}$.

As can be seen in Fig.~\ref{fig:results_random}(2), for small rank values $\mathcal{R} < 8$ the input Hamiltonians have sufficiently many Pauli-Strings to find a symmetry without the ACE being employed (orange and green points overlapping). This is not surprising, as the initial number of Pauli-Strings $N \approx 2\mathcal{R}^2$ is approximately $2^{\mathcal{R}}$ for $\mathcal{R} < 8$. For larger ranks and our design specifications (see Appendix~\ref{appen:mlspecs}), the minimum [average] expanded context ratio grows as $O(2^{0.732\mathcal{R}})$ [$O(2^{0.247\mathcal{R}})$] (dash-dotted lines), orders of magnitude below the upper-bound.

The failure probability $p_{\rm f}$ of the optimizer is shown in Fig.~\ref{fig:results_random}(3) alongside the predicted number of independent starts $n_{\rm s}$ and GPUs $n_{\text{GPU}}$ that are required to find a symmetry with $90\%$ success probability. We distinguish between $n_{\rm s}$ and $n_{\text{GPU}}$ because, albeit related, the latter includes additional physical constraints (see next sections). As expected from the inverse relation between $\mathcal{R}$ and the number of Pauli symmetries (see Sec.~\ref{sec:background}), $p_{\rm f}$ (red dots) grows with the rank $\mathcal{R}$. Consequently, the computational resources $n_{\rm s}$ and $n_{\text{GPU}}$ to find a symmetry also increase. 

As we have seen in Fig.~\ref{fig:results_random}(2), the attention-based optimizer requires more context at higher ranks. This also increases the GPU memory requirements, as memory cost for computing attention grows with the context size \cite{NLP_T_vaswani2017attention, NLP_ST_lee2018transformer}. As we use Flash-attention \cite{Flashattention} in our architecture, this increase is linear. Additionally, parallel processing of multiple independent starts further increases the memory requirements. All of this must be taken into account within $n_{\rm GPU}$. To compute it, we first determine the maximum GPU utilization for a single run $\overline{u_{\text{max}}}$, and define 
$$
n_{\text{GPU}} 
\approx 
\lceil{\text{max}(1, n_{\rm s} \overline{u_{\text{max}}})} \rceil.
$$ 
At high ranks, $n_{\text{GPU}}$ coincides with $n_{\rm s}$, as the optimization process saturates the entire GPU. The specifications of the GPU used in this work can be found in Appendix~\ref{appen:mlspecs}.

On the other hand, to calculate $n_{\rm s}$ we treat each start of the optimization process for a single Hamiltonian $H$ to be as an independent Bernoulli trial, such that
$$
n_{\rm s} 
= 
\left\lceil \frac{\ln(0.1)}{\ln(p_{\rm f})} \right\rceil.
$$ 
As can be seen in the plot, at rank $18$ the predicted $n_{\rm s}$ is $32$. As explained in Sec.~\ref{sec:background}, in the optimization landscape of our problem, the ratio of symmetry to non-symmetries of an $H$ with $\mathcal{R}=18$ is $2^{-18}$. This indicates that the attention-based optimization framework accesses a much larger basin of attraction \cite{BOA} compared to random search.

\subsection{$1$-D periodic transverse Ising model}\label{ssec:Ising}

In this section, we present the performance of our optimization framework against physical models. In Fig.~\ref{fig:results_ising}, we use periodic $1$-D transverse-field Ising Hamiltonians \cite{Ising_chain} as a way to benchmark the CPU (light blue) and GPU (blue) implementations (for device specifications, see Appendix~\ref{appen:mlspecs}) of our attention-based optimizer, and compare the results against the deterministic algorithm of Ref.~\cite{symm_Luca_min_qrep} (grey variations). 

For a periodic chain of $n_{\rm q}$ spins the input for our framework is \cite{Ising_chain}
\begin{equation*}\label{eq:ising}
    H_{\text{Ising}}
    = 
    -J \sum_{i=1}^n{Z_iZ_{i+1}} 
    -h_x \sum_{i=1}^n X_i.
\end{equation*}
This Hamiltonian contains $2n_{\rm q}$ Pauli-Strings under the assumptions of periodic boundary conditions and the transverse field. 

To study the CPU and GPU implementations of the attention-based framework against the deterministic algorithm in Ref.~\cite{symm_Luca_min_qrep}, we vary the number of qubits over a range from $n_{\rm q}=10$ to $n_{\rm q}=1400$. For each $n_{\rm q}$, the optimizer was initialized for $100$ independent starts, and a maximum of $200$ iterations with early-stopping criteria in place (for details, see Appendix~\ref{appen:mlspecs}). The deterministic algorithm in Ref.~\cite{symm_Luca_min_qrep}, that cannot be parallelized, is run on the same CPU as the optimizer.

Fig.~\ref{fig:results_ising}(1) shows the average time taken to find a symmetry. Each datapoint is averaged over the successful runs out of the $100$ starts, with error bars representing standard deviations. Fitting the run-times to a power-law $\beta n_{\rm q}^\alpha + \gamma$ (dashed-dotted lines) yields exponents $\alpha = 2.96$, $\alpha = 2.75$, and $\alpha = 3.44$ for the CPU, GPU, and deterministic implementations, respectively. Within statistical errors, the first two are compatible with each other. Concerning the deterministic approach, $\alpha = 3.44$ is slightly higher than the expected value of $3$ attained in the asymptotic limit $N \gg 2n_{\rm q}$ \cite{symm_Luca_min_qrep} (see Sec.~\ref{ssec:res_rand}). 

It is more interesting to compare the coefficients $\beta$ from the three fits, as they quantify the advantage of our framework. With the exponents $\alpha$ being similar to each other, this advantage remains constant when increasing the qubits. For the CPU, GPU, and deterministic implementations, we find $\beta \approx 3.25 \times 10^{-7} $, $\beta \approx 4.38 \times 10^{-8}$, and $\beta \approx 7.31 \times 10^{-5}$ seconds, respectively. This means that for sufficiently large instances, our framework finds a symmetry $\approx 1500$ [$\approx 225$] times faster than the deterministic approach when employing the GPU [CPU].

In Fig.~\ref{fig:results_ising}(2) we show the iterations required to find a Pauli symmetry and the number of Clifford gates in the diagonalizing circuits. The first is relevant for our framework, while the latter for the deterministic approach -- each gate corresponds to an algorithmic step \cite{symm_Luca_min_qrep}.
For all Hamiltonians considered, our optimizer finds a Pauli symmetry within $100$ iterations across system sizes, requiring similar numbers of iterations for the CPU (dark-blue) and GPU (light-blue) implementations.

Interestingly, the iterations required are higher for smaller system sizes, and saturate at $\approx 35$ for $n_{\rm q} \gtrsim 500$. This behavior is explained by the early stopping conditions, which trigger earlier when $n_{\rm q}$ increases. We expect this behavior to be driven by the interplay between the binary regularizer in Eq.~\eqref{eq:loss_bin} and the enforced zero-penalty in Eq.~\eqref{eq:loss_zp}. For small system sizes and during early training, the first struggles to move away from $S(\theta_i) = 0.5$ as the latter explodes when the sum of output terms is close to zero. 
Concerning the deterministic approach, Fig.~\ref{fig:results_ising}(2) shows that the number of Clifford gates increases with the number of qubits. This is in contrast with the iterations required by the optimizer, which remain approximately constant for both CPU and GPU implementations for increasing $n_{\rm q}$.   

Finally, in Fig.~\ref{fig:results_ising}(3) we show the probability of failure $p_{\rm f}$ (dots) alongside its standard deviations (shadows), computed from the number of unsuccessful optimizations out of the $100$ starts. We find that between system sizes and implementation types (GPU and CPU), the average and minimum $p_{\rm f}$ are $0.033 \pm 0.008$ and $0.009\pm 0.002$, respectively. We note that $p_{\rm f} = 0.01$ indicates that all runs were successful (follows from Bayesian statistics with uniform prior \cite{shlosberg2023adaptive, simon2025enhanced, Bays_pf2, Bays_pf3}). In particular, between $n_{\rm q} = 40$ and $n_{\rm q} = 700$, only $10$ starts out of $13 \cdot 100 \cdot 2 = 2600$ failed.

\subsection{Ising ladder and toric code}\label{ssec:Toric}

In Fig.~\ref{fig:results_toric}, we study the performance of our attention-based optimization framework applied to the periodic $2$-D Ising ladder \cite{Ising_ladder} (dark red) and for Kitaev's Toric code \cite{KITAEV20032, toric2, PhysRevB.79.033109, ferguson2021measurement, chan2024measurement}, with (dark-orange) and without (light-orange) a magnetic field $\vec{B}$. 
All simulations with our optimization framework were implemented on the GPU. For comparison, results from the deterministic algorithm \cite{symm_Luca_min_qrep} are in black (Ising ladder), dark-gray (Toric with $\vec{B}$), and light-gray (Toric without $\vec{B}$).

The Hamiltonian for the $2$-D Ising ladder with $n_y = 2$ legs and $n_x$ rungs ($\implies n_{\rm q} = n_y n_x = 2n_x$) is given by
\begin{equation*}
H_{\rm IL} 
= 
J \sum_{\langle i,j \rangle}
Z_{i}Z_{j} 
+ h \sum_{i=1}^{n_{\rm q}} X_{i}
,
\end{equation*}
where $\langle i,j \rangle$ indicates connected qubits $i$ and $j$.
On the other hand, the Toric code considers a rectangular lattice with periodic boundary conditions on a torus. The Hamiltonian can be written (in presence of a magnetic field $\vec{B}$) as
\begin{equation*}\label{eq:toric}
    H_{\text{Toric}} 
    = 
    c_x \sum_{p} 
    \left( \prod_{e \in p} X_e \right) 
    + c_z \sum_{s} 
    \left(\prod_{e \in s} Z_e \right) 
    + c_g \sum_e Z_{e},
\end{equation*}
where $p$ and $s$ are the plaquettes and stars, respectively \cite{KITAEV20032, toric2, PhysRevB.79.033109, ferguson2021measurement, chan2024measurement}. The term $c_g \sum_e Z_{e}$ is associated to the magnetic field $\vec{B}$; in absence we set $c_g=0$.

Concerning the simulations, we let $n_{\rm q}$ range from $10$ to $1000$ ($100$ for the deterministic approach \cite{symm_Luca_min_qrep} due to time constraints), allowed for $100$ independent starts per Hamiltonian, set the number of maximum iterations to $400$, and enabled early-stopping criteria (see Appendix~\ref{appen:mlspecs}). While the Ising ladder only has one possible configuration for a given $n_{\rm q}$, the Toric code has multiple depending on the torus' sizes chosen. The data points in Fig.~\ref{fig:results_toric} are averaged over random instances with fixed $n_{\rm q}$. Dash-dotted lines are numerical fits.

Similarly to the periodic Ising chain in Fig.~\ref{fig:results_ising}(1), our attention-based optimization framework provides a substantial advantage in the time taken to find the Pauli symmetry of $H_{\rm IL}$ over its deterministic counterpart. We find that both the GPU implementation of the optimizer and the deterministic approach for $2$-D Ising ladder scale approximately as $O(n_{\rm q}^{3.44})$, although the GPU implementation has a multiplicative coefficient that is $10^5$ times smaller compared to the deterministic algorithm ($\approx 10^{-10}$ against $\approx 10^{-5}$ seconds, respectively).

On the other hand, we find that the Toric codes have differing scalings if we compare our framework against the deterministic algorithm. Fits of the first [latter] yield $O(n_{\rm q}^{3.268})$ and $O(n_{\rm q}^{3.107})$ [$O(n_{\rm q}^{4.669})$ and $O(n_{\rm q}^{4.754})$] for the instances with and without magnetic field, respectively. This implies that the larger the torus is, the more advantageous our approach becomes. Similarly to what we have seen in Sec.~\ref{ssec:res_rand}, the deterministic approach does not saturate the scaling $O(n_{\rm q}^3)$. We believe that this follows from the moderate number of Pauli-Strings.

In Fig.~\ref{fig:results_toric}(2), we show the iterations (number of gates $n_{\rm gates}$) required by the attention-based (deterministic) model to find a Pauli symmetry. Similarly to the $1$-D Ising Chain in Fig.~\ref{fig:results_ising}(2), the optimizer requires a decreasing number of iterations to find Pauli symmetries across the considered range of $n_{\rm q}$. It asymptotically stabilizes at approximately $40$, and $60$ iteration for $n_{\rm q} \gtrsim 500$ for the Ising and the two toric models, respectively. In comparison, the number of gates in the Clifford circuit used by the deterministic algorithm to find the Pauli symmetries \cite{symm_Luca_min_qrep} increase with $n_{\rm q}$. Specifically, the growth is more pronounced for the two toric models.

Finally, the failure probabilities $p_{\rm f}$ are presented in Fig.~\ref{fig:results_toric}(3). Across the entire range of qubits considered, the optimization framework achieves a high success rate for all models. We ascribe the larger variations seen for the toric code to varying performances attained for different geometries of the model (see above). However, in all considered cases and for all models, the average and minimum $p_{\rm f}$ are $0.038 \pm 0.005$ and $0.010X \pm 0.002$, respectively. Between $n_{\rm q} = 40$ and $n_{\rm q} = 700$, only $72$ starts out of $13 \cdot 100 \cdot 3 = 3900$ failed.

\subsection{Remarks}

As shown in Fig.~\ref{fig:results}, the attention-based optimizer provides substantial advantage over the deterministic algorithm in Ref.~\cite{symm_Luca_min_qrep} for the considered physical Hamiltonians -- see Secs.~\ref{ssec:Ising} and \ref{ssec:Toric}. In comparison, for the randomly generated Hamiltonians in Sec.~\ref{ssec:res_rand}, it requires larger computational resources. For the physical systems [Figs.~\ref{fig:results_ising}(2) and~\ref{fig:results_toric}(2)], the model often finds a symmetry within the first context-expansion window, without triggering the ACE introduced in Sec.~\ref{sec:framework}. We attribute this to the \textit{structure} present in the tableau representation of the physical Hamiltonians. Unlike randomly generated ones, the constituent Pauli-Strings of a physical Hamiltonian are not arbitrary, as they encode the ordered physical interactions. As an example, the $1$-D periodic Ising-chain Hamiltonian contains $n_{\rm q}$ Pauli-Strings representing the nearest-neighbor interactions $Z_iZ_{i+1}$, and $n_{\rm q}$ terms encoding the transverse field along $x$. This structure maintains locality ($2$-local in case of Ising chain), repetitive motifs (only nearest-neighbors or single qubit Pauli), as well as sparsity in data. This makes the context available to the optimizer immediately informative. As a result, the optimizer navigates the complex landscape of the commutation loss more easily. In contrast, the randomly generated Hamiltonians suffer from the absence of this regularity, and need more resources (in terms of both context and time) to detect the global relation (i.e., commutation) encoded by the symmetry.

\section{Conclusion}\label{sec:conc}
We developed an optimization framework with an underlying Set-Transformer architecture to find the Pauli symmetries of a Hamiltonian. The optimizer is GPU-enabled and allows for parallelization of the symmetry-finding process. We applied our model to find a Pauli symmetry of randomly generated Hamiltonians as well as the $1$-D periodic transverse-field Ising Chain, the $2$-D Ising Ladder and the Toric code. We find that the attention-based optimizer works particularly well with these physical Hamiltonians, where systemic physical interactions are embedded into their tableau form. Specifically, the GPU-enabled optimizations found symmetries much faster than existing approaches \cite{symm_Luca_min_qrep, symm_taperingqubits, van_den_Berg_2020circuit}, and with a high success rate. For randomly generated Hamiltonians, on the other hand, we need large computational resources (GPUs to facilitate parallelized independent starts) to find symmetries faster than the existing approaches. 

Our framework is, to our knowledge, the first approach to merge machine learning with automated symmetry finding for arbitrary input Hamiltonians. It can already advance the fields of classical and quantum simulations of physical systems. To reach full potential, we plan to extend our approach to find other symmetry classes of Hamiltonians (e.g., Clifford \cite{symm_PRL, symm_companion}), where no optimal strategy is known. 

\section*{}
\bibliographystyle{apsrev4-1}
\bibliography{references}

\appendix
\section{Loss Functions}\label{appen:loss}
The parameters of the Set-Transformer model is optimized primarily using a custom loss function that leverages the commutation relation between a Hamiltonian $H$ and its Pauli symmetries. The output of the decoder stack (see Sec.~\ref{sec:framework}), $S(\theta)$ represents the tableau representation of a Pauli-String, dependent on the system parameters $\theta$. As all Pauli-Strings $P_i$ of an input Hamiltonian $H$ commute with any of its Pauli symmetries $S_{\rm p}$, the symplectic inner product of any $P_i$ and $S_{\rm p}$ (modulo $2$) is zero. We use this to construct our primary loss function for the optimizer, i.e., the commutation loss as (see Sec.~\ref{sec:background}; Eq.~\eqref{eq:loss_com}),

\begin{equation*}\label{commlos}
    {\mathcal{C}_{\text{com}}(\theta)} = \sum_{k=1}^{N}\sin^2\left(\frac{\pi}{2}*[ (H_{\rm t}J)_kS(\theta)^{T}]\right),
    \end{equation*}
where $H_{\rm t}$ is the tableau representation of the Hamiltonian $H$ and $J$ is the canonical symplectic form on the vector space $\mathds{F}_2^{2n}$ \cite{sympleq_Haah_2017}. 

To maintain the differentiability of the commutation loss we use a differentiable proxy of modulo $2$, as $Sin^2(*)$. Ideally, if the optimizer finds a Pauli symmetry of $H_{\rm t}$, the commutation loss becomes $0$. Our model minimizes the function $\mathcal{C}_{\text{com}}(\theta)$ to obtain the same. 

$\mathcal{C}_{\text{com}}(\theta)$ is our main loss function. However, we add three helper penalties to aid the optimization process. The first penalty prevents the optimizer to find the Identity, the trivial Pauli symmetry of every Hamiltonian. As the Identity is represented by a $2n_{\rm q}$-dimensional vector of all $0$s in its tableau form, the penalty is designed as (see Sec.~\ref{sec:background}, Eq.~\eqref{eq:loss_zp}), 
\[\mathcal{C}_{\text{zp}} = \left(\sum_{i=1}^{2n_{\rm q}}S(\theta)_i \right)^{-2},
\]
where $S(\theta)_i$s are the elements of the output vector $S(\theta)$. 

Next, a binary regularizer is added that penalizes the values of $S(\theta)_i$ close to $0.5$. As the Set-Transformer uses continuous values between $0$ and $1$ for the elements of $S(\theta)$ during training while trying to predict a binary vector, this penalty ensures that $S(\theta)_i$s get closer to either $0$ or $1$ as optimization progresses. The regularizer is designed as (Sec.~\ref{sec:background}, Eq.~\eqref{eq:loss_bin})),  
\[ \mathcal{C}_{\text{bin}} = \sum_{i=1}^{2n_{\rm q}}S(\theta)_i(1-S(\theta)_i).
\]

As the landscape of the $Sin^2(\cdot)$ function in the commutation loss is highly multi-modal, optimization becomes costly, sensitive to initial values, and prone to struggle near local minima. To help the optimizer negotiate this non-linear landscape faster, we have added a linearity-regularizer  (Sec.~\ref{sec:background}, Eq.~\eqref{eq:loss_lin})):
\[\mathcal{C}_{\text{lin}} =\sum_k \left\lvert (H_{\rm t}J)_k S(\theta)^{T}
- \text{ne}\left(\sum_{l} (H_{\rm t}J)_{k,l}\right)\right\rvert.
\]
The $(H_{\rm t}J)_k S(\theta)^{T}$ term computes a continuous-valued symplectic inner product between the $k$th Pauli term in the Hamiltonian and the predicted symmetry. If $S(\theta)$ is indeed a symmetry of $H$, this product should be an even integer. The regularizer aims to minimize the difference between this product and a reference term, the nearest even integer approximation of the sum of the $k$th row of $H_{\rm t}J$. While this regularizer does not guarantee a minimum value at a symmetry, it is piece-wise linear, and helps stabilizing the training in the initial phase. We use the linearity regularizer with a initial weight based on the ratio of the initial values of itself and the primary loss function $\mathcal{C}_{\text{com}}(\theta)$, and its contribution is then gradually reduced over the course of training through a ramp-down schedule.

\section{Adaptive Context Expansion Algorithm}\label{appen:ace}

\begin{algorithm}[t]
\caption{Adaptive Context Expansion}
\label{alg:ace}
\KwIn{Current iteration $t$, update step $t_{\mathrm{upd}}$, updation window size $W$, current loss $\ell(t)$, decision window size $r < W$, tolerance $\epsilon$}
\KwOut{ExpandContext flag}

\If{$t = t_{\mathrm{upd}} + 1$}{
    $\ell_{\min} \gets \ell(t)$\;
    
}

\For{$i \gets 1$ \KwTo $W$}{
    $\ell_{\min}(i) \gets \min\{\ell_{\min}(i-1),\, \ell(t)\}$\;
    $\Delta_W(i) \gets \ell_{\min}(i-1) - \ell_{\min}(i)$\;
}

\For{$j \gets W-r$ \KwTo $W$}{
    $\Delta_{\mathrm{r}}(j) \gets \ell_{\min}(j-1) - \ell_{\min}(j)$\;
}

\If{$t = t_{\mathrm{upd}}$}{
    \If{$\overline{\Delta_{\mathrm{r}}} <\lvert\overline{\Delta_{W}} - \sigma(\Delta_W)\rvert +\epsilon$}{
        \textsc{ExpandContext}\;
    }
}
\end{algorithm}

 We introduce our algorithm to dynamically expand the context size available to the Set-Transformer. If two Pauli-Strings $P_i$ and $P_j$ commute individually with a third Pauli-String $P_k$, their symplectic sum $P_i + P_j$ ($+ :=$ addition modulo $2$) also commutes with $P_k$. By adding the symplectic sum of two arbitrarily chosen Pauli-Strings in the Hamiltonian, the context available to the model can be increased, while keeping it relevant to the optimization objective, i.e., finding the Pauli symmetry. 

The context-expansion algorithm is adaptive to the performance of the model. At fixed iteration intervals, a context expansion window of size $W$ is initiated. The minimum loss throughout this window is recorded, and its stepwise decrement is computed and stored as $\Delta_{W}$. A fraction of $W$ is then selected as the decision window, which consists of the last $\mathcal{R}$ steps of $W$. The algorithm then compares the stepwise average decrement of minimum loss for the decision window, $\overline{\Delta_r}$, with $\overline{\Delta_W}$. We infer that the optimization progress is struggling if $\overline{\Delta_r} < \lvert\overline{\Delta_W} - \sigma(\Delta_W)\rvert +\epsilon$, where $\sigma(\Delta_W)$ is the standard deviation of $\Delta_W$, and $\epsilon$ is a small positive tolerance. This indicates the optimizer is struggling to make consistent progress. It is either stuck near a local minimum, or exhibiting noisy progress over the expansion window (large $\sigma(\Delta_W)$ compared to $\overline{\Delta_W}$), while the average recent improvement ($\overline{\Delta_r}$) remains small. In either case, the algorithm expands the available context. The size of the expansion is design dependent. We use the absolute value of $\overline{\Delta_W} - \sigma(\Delta_W)$ to ensure a non-negative expansion threshold, whereas the small positive tolerance $\epsilon$ ensures that an expansion triggers in case of near-zero improvements. We provide a pseudocode of the module in Algorithm~\ref{alg:ace}.

\section{Device and design  Specifications}\label{appen:mlspecs}

\begin{table*}[t]
    \centering
\captionsetup{justification=centering}
     \caption{Design Specifications.}
 \label{tab:mlspecs}
 \renewcommand{\arraystretch}{1.3}
 \begin{tabular}{|l|l|}
    \hline
    \textbf{Conditions} & \textbf{Details}\\
    \hline
    Initiation  &T-fixup Schedule (Encode-Decoder layers) \cite{NLP_tfixup}.\\
                  & Xavier (Linear Layer).\\
                  & Constant (Layer Norm).\\ 
                  & Bias: 0.\\
    \hline 
    Multi-Start & Yes (Independent starts).\\
   \hline 
    Optimizer  &  AdamW (Betas: $0.9$, $0.99$).\\
    \hline
    Iterations & Fixed maximum value (with early stopping).\\
    \hline
    Learning Rate & Linear warm-up with Cosine decay.\\
                  & (Highest value: 1e-3, Warm-up: Maximum Iteration/$3$).\\
    \hline
    Gradient Clipping & Yes (Maximum norm: $1$).\\ 
    \hline
    Early Stopping Conditions 
                   & 1. Output ($S(\theta)$) commutes with input ($H$) [$HJS(\theta)= 0 \pmod{2}$].\\
                   & 2. At least one element in $S(\theta)$ is nonzero.\\
                   & 3. $C_{\rm bin}(\theta)= \sum_{i=1}^{2n_{\rm q}}S(\theta)_i(1-S(\theta)_i) > \text{fixed cutoff} \approx \frac{n_{\rm q}}{10}$.;\\ 
                   & $n_{\rm q}$: number of qubits, $C_{\rm bin}(\theta)$: binary regularization.\\
                   & \textit{note}: All three conditions needs to satisfied.\\ 
   \hline
    
    \end{tabular}
   
\end{table*}
All simulations are performed with a NVIDIA RTX 2000 Ada Generation (GPU) (16\,GB VRAM; compute capability 8.9), or a Intel Core i7-10700 CPU (8 cores/16 threads, 2.90 GHz) and 32 GB RAM. The deterministic algorithm was implemented using the SympleQ \cite{sympleq} using the same CPU. 

We provide our design specifications in Table \ref{tab:mlspecs}. The model and optimization routines were implemented using PyTorch \cite{pytorch}. An open-source implementation of the attention-based optimizer can be found in Ref.~\cite{sympleq}.

\end{document}